\documentclass{article}
\usepackage{amssymb}
\usepackage{epsfig}
\usepackage{amsmath}
\usepackage{times}
\usepackage{ulem}
\usepackage{color}
\usepackage{graphicx}
\usepackage{a4wide}
\graphicspath{{}{../matlab/resultfig/}{figures/}}

\newcommand{\s}{\sigma}
\newcommand{\la}{\lambda}

\newcommand{\nn}{\nonumber}

\newcommand{\bsub}{\begin{subequations}}
\newcommand{\esub}{\end{subequations}}

\newcommand{\vci}[1]{\begingroup
\setbox0=\hbox{{#1}}%
\parbox{\wd0}{\box0}\endgroup}

\begin{document}

\title{
Inversion of electromagnetic induction data\\  using a 2D geophysical response function
}
\author{Hans Dierckx$^1$, Katrien De Blauwe$^1$, Marc Van Meirvenne$^2$, Henri Verschelde$^1$}
\date{ $^1$ Department of Physics and Astronomy, Ghent University, Gent, Belgium\\
$^2$ Department Environment, Ghent University, Gent, Belgium\\
}

\maketitle

\abstract{Electromagnetic induction methods are a common means for geophysical survey. For soil structures that are invariant in one spatial dimension such as trench structures, we propose a fast forward model based on a 2D response function, taking into account sharp horizontal and vertical transitions in the electrical conductivity. The 2D model is used to invert trench structures from a synthetic dataset. The propsed method can determine trench parameters (width, height, steepness of the slope) more accurately than methods based on regularization of the 1D response function.}


\section{Introduction}

The basic principles of electromagnetic induction surveying are simple: a transmitting coil generates a primary electromagnetic field that interacts with the environment. As a result, the receiving sensors in the instrument will register a secondary field that is determined by the environment. In frequency-domain instruments, the amplitude of the quadrature phase response in the receiver coil is commonly expressed in units of apparent electrical conductivity (EC). The apparent electrical conductivity (EC) is influenced by many factors, and has therefore been used to map organic soil content \cite{huang:2017}, salinity \cite{dakak:2017}, and underground conducting bodies such as unexploded ordnance (UXO) \cite{Grzegorczyk:2012, oneill:2016} or mineral ores \cite{Vallee:2011}. As such, applications of electromagnetic induction (EMI) surveys include (precision) agriculture \cite{huang:2017, dakak:2017}, mineral exploration \cite{Vallee:2011}, soil remediation \cite{Beard:1998}, as well as archaeology \cite{DeSmedt:2014, Saey:2014}. Depending on the application, sensors can be pushed, towed, airborne or mounted on underwater vehicles. 

In this work we focus on the reconstruction of a multi-layered conducting earth, where lateral discontinuities between the layers with different electrical conductivity are present. 

The general solution for a horizontally stratified earth was first given by Wait \cite{Wait:1962} using a transfer matrix formalism. In the approximation that different horizontal layers respond independently (i.e. at low induction number), McNeill \cite{McNeill:1980} and Wait \cite{Wait:1962} derived simple cumulative response functions $R_c(z)$ for different orientations of the transmitter and receiver coils:
\begin{align}
 R_{HCP}(\tilde{z}) &= \frac{1}{(4\tilde{z}^2+1)^{1/2}}, \\
 R_{PERP}(\tilde{z}) &=1- \frac{2\tilde{z} }{(4\tilde{z}^2+1)^{1/2}}. 
\end{align}
Here $\tilde{z} = z / s$ with $z$ the depth relative to the sensor and $s$ the distance between transmitter and receiver coils. 
In the1D forward model, the apparent electrical conductivity is given by
\begin{align}
   \sigma_{app, c} &= \sigma_0 R_c(z_0) + \sum_{j=1}^N (\sigma_j - \sigma_{j-1}) [ R_c(z_j) - R_c(z_{j-1})] \label{MNfwd}
\end{align}
with $z_j$ the depth of the interface between the $j$-th and $(j+1)$-th layer that has conductivity $\sigma_j$.  
Recently, damping effects in moderately to highly conductive soils have been added to the description \cite{Delrue:2018}. 

The parameters $p_i$ in Eq. \eqref{MNfwd} to be estimated in a survey are the layer depth $z_j$ for each measurement location and layer conductivity $\sigma_j$, which may be taken constant or slowly varying. A discretised forward model is generally a function converting parameters $p_i$ into estimates of observables $m_k$:
\begin{align}
   \hat{\mathbf{m}} =  \mathbf{F} ( \mathbf{p}). 
\end{align}
In the model \eqref{MNfwd}, the mapping is performed for every measured location independently. It is often called a 1D model since the $R_c$ depends on a single spatial coordinate $z$.  

To reconstruct geophysical structures from EMI measurements, an inversion step is needed. Hereto, one typically constructs a forward model for the electromagnetic response of the subsoil feature or object (e.g. interface between layers or an infill or object), and thereafter minimises the mismatch $E$ between the measured data $m_k$ and the data $\hat{m}_k $ as predicted by the forward model given model parameters $p_i$:
\begin{align}
  E[p_i] = \sum_{\rm{measurements}\ k}    d( m_k , \hat{m}_k(p_i) ) \quad  + P( p_i) \label{cost}
\end{align}
Here, $d(x,y)$ is a distance function. E.g. $d(x,y) = (x-y)^2$ produces a least-squares method. Since there are usually more unknowns than observations, it is in practice necessary to further constrain the solution by adding a penalty term $G$ that disfavours non-smooth or non-physical (e.g. large-amplitude) solutions. 
The best-known example in this context is Tikhonov regularisation \cite{Tikhonov:1977}, in which $P = \lambda \mathbf{p}^T \boldsymbol{\Gamma}^T \boldsymbol{\Gamma} \mathbf{p}$. A common choice is to take for  $\Gamma$ an implementation of the spatial Laplacian operator. In most cases, $E$ is a non-linear function of $\mathbf{p}$ and a non-linear solver must be used to find the set of model parameters $\mathbf{p}$ that minimises the cost function $E$. 

In the context of EM inversion, the laterally constrained inversion method (LCI) implements a penalising term $G$ that lowers the mismatch of model parameters between adjacent measurement points \cite{auken:2004, MonteiroSantos:2004, MonteiroSantos:2010}. The method is often referred to as 2D or 3D inversion, although the underlying response functions $R_c(z)$ still depend on the z-coordinate only. 

Another way of regularising the inversion is to combine multiple data obtained at one measurement location (e.g. using different frequencies or coil position and orientation) and perform a single joint inversion \cite{farquharson:2003, brodie:2009, Gholami:2010}. 

To handle complex subsurface structures, mesh-based forward models that are resolved in 2 or 3D spatial dimensions have been used \cite{sasaki:2001, sasaki:2010, noh:2016}. These forward models are computationally expensive, and inversion may take many hours to compute, even on distributed architectures. For this reason, different accelerating techniques are being investigated, at the level of the model, inversion algorithm or non-linear solver \cite{thiesson:2017}. 

In this work we take an approach that lies in between the classical 1D and fully resolved 2D or 3D forward models. In the cases where the features of a multi-layered earth vary rapidly only along one horizontal spatial dimension (say, $x$) the 2D response function $R(x,z)$ will be computed analytically and thereafter used for inversion. 

This paper is organised as follows. In Section \ref{sec:methods} the forward model and inversion routine is described. In Section \ref{sec:inversion} we study the performance of the method in the reconstruction of a steep 2-sided trench with different electrical conductivity.  In Sec. \ref{sec:discussion} we discuss future applications of our method. 


\section{Materials and Methods \label{sec:methods}}

\subsection{Forward modeling}

\subsubsection{Dipole models of transmitter and receiver coils}

The measurement principle of a geophysical induction sensor is to emit radiofrequent waves using a transmitting coil $T$, which induce Eddy currents in the soil due to its low but finite conductivity. These Eddy currents generate a secondary magnetic field $\vec{H}_s$ which is sensed by the receiver coils $R$ of the sensor.

We will indicate the coil orientation in terms of its symmetry axis $\vec{e}_m$ or magnetic moment $\vec{m} = m \vec{e}_m$. We will write $\vec{m}_1 = m_1 \vec{e}_1$ for the transmitting coil and $\vec{m}_2 = m_2 \vec{e_2}$ for the receiver. Likewise, $\vec{r}_1$ and $\vec{r}_2$ will denote the vectors pointing from the center of $T$ and $R$ towards a given point; see Fig. \ref{fig:config}. Furthermore, $r_1 = ||\vec{r_1}||$, $r_2 = || \vec{r_2}||$. 

To fix thoughts, we here model measurements made with a DUALEM-21S instrument (DUALEM, Milton, Canada) which has a single transmitter coil with vertical axis: $\vec{m}_1 = m_1 \vec{e}_z$. At a horizontal distance $- \vec{s} = -  s \vec{e}_x$ from the transmitter, a receiver coil is placed, which can be either in the `horizontal co-planar orientation' (HCP, $\vec{m}_2 = m_2 \vec{e}_z)$ or `perpendicular' orientation (PERP, $\vec{m}_2 = m_2 \vec{e}_x)$. The Dualem-21S has two HCP coils at $s = 1$ and $2$ meter, and two PERP coils at $s=1.1$ and $s=2.1$ meter, respectively. 

We will only use the quadrature signal from the receiver coils, i.e. $H_{s,q} =Im(\vec{H}_s \cdot \vec{e_2})$.
 In the sensor hardware, the signal strength is converted signal to an apparent electrical conductivity using \cite{McNeill:1980}
\begin{align}
   \s_{app} = \frac{4}{\omega \mu s^2} \frac{H_{s,q}}{H_{p,ref}} \label{sapp0}
\end{align}
where $H_{p,ref}$ is the in-phase response that would be experienced by a coil at the receiver position that is parallel to the transmitting coil (i.e. maximally coupled). 

Modelling the emitter as a magnetic dipole, we have that 
\begin{align}
  \vec{H}_p = \frac{1}{4\pi} \left( \frac{3 \vec{r}_1 (\vec{m}_1\cdot \vec{r}_1) }{r_1^5} - \frac{\vec{m}_1}{r_1^3} \right)
\end{align}
such that in case of maximal coupling with intercoil distance $s$, one has
\begin{align}
  H_{p,ref} = \frac{m_1}{4 \pi s^3}. \label{Hpref}
\end{align}
Combined with Eq. \eqref{sapp0}, one finds:
\begin{align}
   \s_{app} = \frac{16 \pi s }{\omega \mu } H_{s,q}. 
\label{sapp}
\end{align}

\subsubsection{Green's function approach}

We now build a forward model for the generation of the signal in the EMI sensor above a non-metallic, non-magnetic soil ($\mu \approx \mu_0$) with arbitrary electrical conductivity $\s(\vec{r})$. We work with a low induction number (LIN-approximation), which allows to neglect feedback loops and mutual induction between the transmitter and receiver coils. Our analysis is performed in the frequency domain, where $\omega = 2\pi f$, $f =9$kHz is the frequency of the signal generated in the transmitter coil. 

\begin{figure} \centering
 \includegraphics[width=0.4\textwidth]{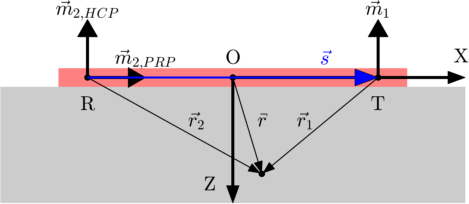}
\caption{\label{fig:config} Coordinate system for calculations, showing the relative position and orientation of the transmitting (T) and receiving coils (R) in the EMI-sensor (red).   
}

\end{figure}

This process is modeled as follows; see Fig. \ref{fig:config} for a sketch of the set-up. We take the origin of our coordinate system at the centre of a transmitter-receiver pair, with the transmitter at $\vec{s}/2$ and receiver at $-\vec{s}/2$. The distance to a point $\vec{r}$ where an eddy current is generated is $\vec{r_1} = \vec{r} - \vec{s}/2$ from the transmitter and $\vec{r_2} = \vec{r} + \vec{s}/2$ from the receiver. 

The transmitter coil produces a primary magnetic field $\vec{H}_p$ which introduces electric current densities $\vec{J}(\vec{r})$ in the soil. Approximating the transmitter coil by an oscillating magnetic dipole with magnetic moment $\vec{m}_1 = m_1 \vec{e_1}$ oriented along the coil axis, this source generates a field with vector potential \cite{Jackson:1999}
\begin{align}
\vec{A} = \frac{\mu}{4\pi} \vec{m}_1 \times\frac{ \vec{r_1}}{ r_1^3}
\end{align}
where $\mu$ is the magnetic permeability of the medium, which we take constant and equal to the vacuum value $\mu_0 = 4 \pi. 10^{-7} N/A^2$. 
The corresponding electric field can be expressed in terms of the scalar and vector potential as  
\begin{align}\label{eqElectField}
	\vec{E} = -\nabla U - \frac{\partial\vec{A}}{\partial t}.
\end{align}
It is convenient to work in the temporal gauge (or Gibbs gauge), i.e. to take $U=0$ without loss of generality. Furthermore, in the frequency domain, all time dependence is assumed to be a factor $e^{-i\omega t}$. Then, Eq. \eqref{eqElectField} simplifies to
\begin{align}
\vec{E} = i \omega \vec{A}.
\end{align}
The local electrical field induces eddy currents in the soil, which we approximate by Ohm's law:
\begin{align}
\vec{J} =\s \vec{E} 
\end{align}
Finally, these electric currents generate a secondary magnetic field $\vec{H}_s$ at the receiver, given by Biot-Savart's law 
\begin{align}\label{eqbiotsavart}
 \vec{H}_s = \frac{1}{4\pi} \int \textrm{d}^3 r \vec{J}(\vec{r})\times\frac{-\vec{r}_2}{r_2^3}.
\end{align}
Combining Eqs. \eqref{sapp} - \eqref{eqbiotsavart}, we find that the apparent conductivity readout from the sensor is 
\begin{align}
\s_{app} = \int \s(\vec{r}) G(\vec{r})\ \mathrm{d}^3 r \label{Green}
\end{align}
with Green's function
\begin{align}
G(\vec{r}) &= \frac{s}{\pi} \left( \vec{e}_2 \times\frac{ \vec{r_2}}{ r_2^3} \right) \cdot 
\left( \vec{e}_1 \times\frac{ \vec{r_1}}{ r_1^3} \right)   \label{Greenvector}\\
&= \frac{s}{\pi} \frac{ (\vec{e}_1 \cdot \vec{e}_2)( \vec{r}_1 \cdot \vec{r}_2 ) -  (\vec{e}_2\cdot \vec{r}_1)( \vec{e}_1\cdot \vec{r}_2 ) }{r_1^3 r_2^3}. \nn 
\end{align}

Note that the electromagnetic reciprocity of the problem is manifest here, as the symmetry of the Green's function under exchange of the labels 1 and  2. 
In what follows, it will be advantageous to normalize distances with respect to the intercoil distance $s$. Therefore we define $\tilde{x} = x/s$, $\tilde{y} = y/s$, $\tilde{z} = z/s$ and similar for other coordinates introduced below. By using dimensionless coordinates $\tilde{x} = x/s$ etc. we can make use of lookup tables for fast evaluation of the integral \eqref{Green} in the forward model. 

Introducing the 3D response function
\begin{align}
 \phi^{3D}(\tilde{x}, \tilde{y}, \tilde{z}) =  G(x,y,z)
\end{align}
the HCP and PERP coil configurations have, from Eq. \eqref{Greenvector}, 
\bsub \label{phi3}
\begin{align}
 &\phi^{3D}_{HCP}(\tilde{x},\tilde{y},\tilde{z}) 
=
 \frac{1}{\pi}  \frac{\tilde{x}^2+\tilde{y}^2 - 1/4}{[(\tilde{x}+ \frac{1}{2})^2 +\tilde{y}^2+\tilde{z}^2]^{3/2}[(\tilde{x}- \frac{1}{2})^2 +\tilde{y}^2+\tilde{z}^2]^{3/2}},  \\
& \phi^{3D}_{PERP}(\tilde{x},\tilde{y},\tilde{z}) 
=
 \frac{1}{\pi} \frac{\tilde{z} (1/2-\tilde{x} )}{[(\tilde{x}+ \frac{1}{2})^2 +\tilde{y}^2+\tilde{z}^2]^{3/2}[(\tilde{x}- \frac{1}{2})^2 +\tilde{y}^2+\tilde{z}^2]^{3/2}}.
\end{align}\esub

\subsubsection{Coil sensitivity in the horizontal direction}

\begin{figure}
 \includegraphics[width=0.5\textwidth]{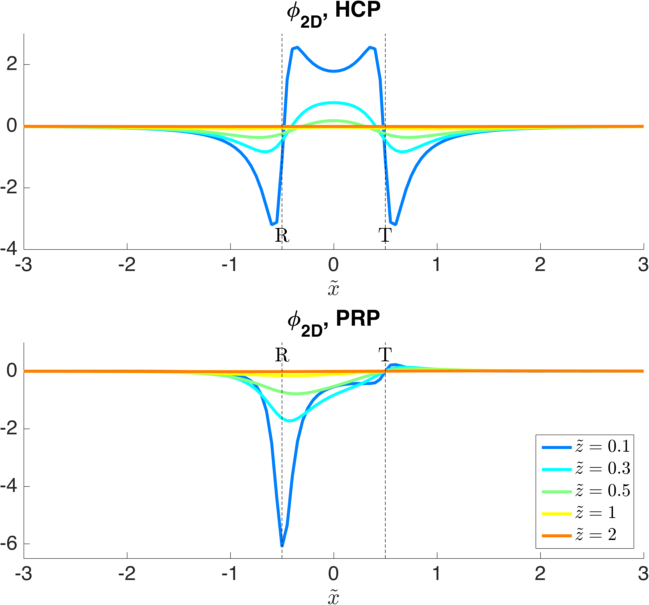}
 \includegraphics[width=0.5\textwidth]{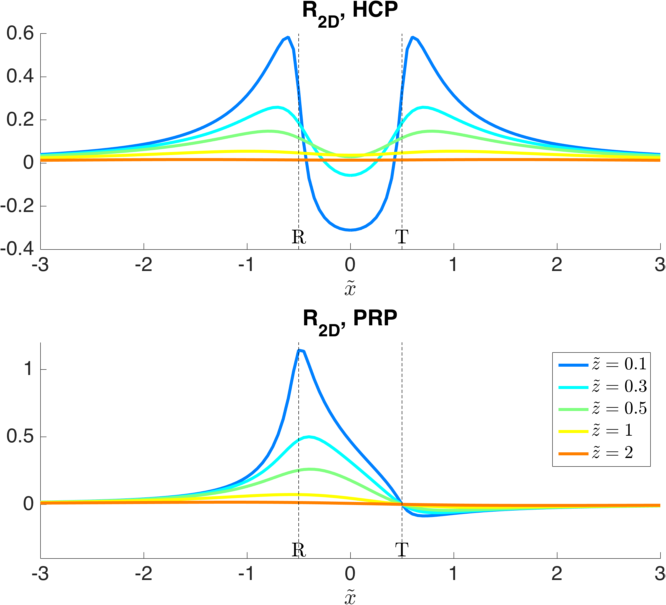}
\caption{\label{fig:R2D} 2D Layer-response curves $\phi_{2D}$ and cumulative depth response curves $R_{2D}$ for receiver coils (R) in HCP and PERP configuration at various relative depths $\tilde{z}$, as a function of horizontal distance $\tilde{x}$. All distances have bee normalised with respect to intercoil distance. }
\end{figure}


In this study, we investigate soil inhomogeneities which slowly fluctuate in one direction, which we assume to be along the y-axis, i.e. during the measurement, one needs to scan along lines perpendicular to the trench structure. 
 When $\s(x,y,z)$ locally behaves like $\s(x,z)$, the integral over $y$ in \eqref{Green} can be carried out explicitly. The subsequent analysis is then performed in terms of the 2D depth response curve and cumulated depth response: 
\begin{align}
  \phi^{2D}(\tilde{x},\tilde{z}) &= \int_{-\infty}^{+\infty} d\tilde{y} \phi^{3D}(\tilde{x},\tilde{y},\tilde{z}), \\
 R^{2D}(\tilde{x},\tilde{z}) &= \int_{\tilde{z}}^{+\infty} d\tilde{z} \phi^{2D}(\tilde{x},\tilde{z}).\label{2D}
\end{align}
The integrals for functions $\phi^{2D}$ and $R^{2D}$ can be evaluated analytically in terms of elliptic functions \cite{Gradstein:1965}. Since the elliptic functions are time-consuming to evaluate during the iterated inversion in Sec. \ref{sec:inversion}, we perform numerical integration to create a lookup table for these functions instead. The resulting curves are plotted for various relative depths $\tilde{z}=z/s$ in Fig. \ref{fig:R2D}. From the exact analytical solution, it is seen that $\phi_{2D}$ becomes discontinuous at finite value at the coil positions $\tilde{x} = \pm 1/2$ in the limit $\tilde{z} \rightarrow 0$. However, this limit case never happens in practice due to the finite radius of the sensor boom. 

Noteworthily, for the HCP coil configuration, objects under both coil positions contribute significantly to the signal, with opposite sign of the object beneath the sensor centre. Since the second order derivative of a function $f$ can be approximated by $f''(x) \approx (f(-h) - 2f(0) + f(h))/h^2$, one infers that a HCP sensor measures the second order spatial derivative of the underlying conductivity in the direction of the sensor axis:
\begin{align}
  \s_{app, HCP} \propto \frac{d^2\s}{dx^2}.
\end{align}

Fig. \ref{fig:R2D} confirms that the PERP receiver coil is most sensitive at shallow depths {\color{red} \cite{elsewhere}}. Moreover, the sensitivity of a PRP coil pair is concentrated under the receiver coil, i.e. 
\begin{align}
  \s_{app, PERP} \propto \s_R.
\end{align}

These observations show that the usual point-wise inversion as proposed in \cite{McNeill:1980} will not accurately reconstruct sudden horizontal variations of soil conductivity. Instead, those methods are designed for soil layers which fluctuate slowly in both $x$ and $y$ directions, in which case one can further integrate Eqs. \eqref{2D} of $x$ to obtain
\begin{align}
  \phi(\tilde{z}) &= \int_{-\infty}^{+\infty} d\tilde{x} \phi^{2D}(\tilde{x},\tilde{z}), \\
  R(\tilde{z}) &= \int_{\tilde{z}}^{+\infty} d\tilde{z} \phi(\tilde{z}) = \int_{-\infty}^{+\infty} d\tilde{x} R^{2D}(\tilde{x},\tilde{z}). 
\end{align}

By changing to cylindrical coordinates $\tilde{x} = r \sin \theta$, $\tilde{y}=r\sin \theta, \tilde{z}=\tilde{z}$, the integrals of Eq. \eqref{phi3} can be explicitly computed, yielding \cite{Wait:1962, McNeill:1980}
\begin{align}
 \phi_{HCP}(\tilde{z}) &=  - \frac{4 \tilde{z}}{(4\tilde{z}^2+1)^{3/2}}, \\
 \phi_{PERP}(\tilde{z}) &= - \frac{2}{(4\tilde{z}^2+1)^{3/2}}, \\
 R_{HCP}(\tilde{z}) &= \frac{1}{(4\tilde{z}^2+1)^{1/2}}, \\
 R_{PERP}(\tilde{z}) &=1- \frac{2\tilde{z} }{(4\tilde{z}^2+1)^{1/2}}. 
\end{align}
For reference, $\phi_V(z), R_V(z)$ from \cite{McNeill:1980} is here denoted as $\phi_{HCP}(\tilde{z})$ and $R_{HCP}(\tilde{z})$; $R(z)$ from \cite{Wait:1962} is written here as $R_{PERP}(\tilde{z})$. 
Since $R({\tilde{z}})$ measures the relative contribution of conductivities deeper than $z$, we have that $R(0)=1$. Note that the formulas in \cite{Saey:2014} describe the cumulative response $1-R$ instead.

\subsubsection{2D layered earth model}
To reach a forward model that is easy enough to invert, we first suppose that the soil consists of $N$ layers of constant conductivity $\s_0,\s_1, ..., \s_{N_1}$. The (j+1)-th layer has constant conductivity $\s_j$ and is bounded by the surface $z=z_j(x,y)$ from above, where $j=0,1,...,N-1$. The top layer ends at a distance $z_0 = 0.16$m from the sensor's central axis:
\begin{align}
 \s(x,z) = \begin{cases}
 0 & \quad \quad \ z < z_0 \\
 \s_j & z_j < z < z_{j+1} \\
\s_{N} & z_N < z
\end{cases}
\end{align}

We restrict ourselves to case where the conductivities are constant in the direction $y$ perpendicular to the measurement direction. Then, from Eq. \eqref{Green}, the apparent conductivity predicted by the multi-layer model is given by
\begin{align}
  \hat{\s}_{app}(x)= \s_0 R(\tilde{z}_0)  + \sum_{j=1}^{N} (\s_{j} - \s_{j-1}) 
 \int_{-\infty}^{+\infty} d\tilde{x}' R^{2D}(\tilde{x}', \tilde{z}_{j}(x+x')). \label{2Dfwd}
\end{align}
This 2D expression reduces to the 1D forward model \eqref{MNfwd} when the interface depth $z_j(x)$ varies only on spatial scales larger than the spatial extent of $R^{2D}$. From Fig. \eqref{fig:R2D}, this distance is estimated to be $3s$. 


\subsection{Inversion }

\subsubsection{Inversion method}
We implemented a simple numerical inversion algorithm based on non-linear Tikhonov regularization \cite{Tikhonov:1977}. The inversion is designed to reconstruct a continuous depth profile $z(x)$, but we describe its discretized version here, which was implemented numerically. 

As the unknowns in the inversion, we take the depths $z_j(x_r) $ of the interfaces between layers at $N_r$ uniformly sampled reconstruction positions $x_r$ along the line of measurement. The conductivity values of the layers can be added to the set of unknowns as a column matrix $\boldsymbol{\s} = (\s_1, \s_2, ... \s_N)^T$.

From the $z_{r} := z_j(x_r)$ and $\boldsymbol{\s}$, the forward models \eqref{2Dfwd}, \eqref{MNfwd} can be used to 
calculate $\hat{\s}_{app}(x_m)$, which predicts the observations $\sigma_{app, m} := \sigma_{app}(x_m)$ made at the set $\{x_m \}$ of $N_m$ points for which measurements are available. This concludes the discretised forward model.

Inversion is implemented by iterative minimisation of the cost function
\begin{align}
 E =& \sum_{c=1}^4 \sum_{m=1}^{N_m} |\hat{\s}^c_{app,m} - \s^c_{app,m} |^2 \label{cost} \\  \nn 
& + \lambda_1 \sum_{j=0}^{N-1} \sum_{r=1}^{N_r-1} | z_{r+1}-z_r |^2
+ \lambda_2 \sum_{j=0}^{N-1} \sum_{r=1}^{N_r-1} | z_{r+1}-z_r |^1. 
\end{align}
Here, $c$ labels the receiver coil. We include two regularisation terms, with relative weights $\la_1$, $\la_2$. The first term is classical Tikhonov regularisation, penalising the gradient of the solution in the L-2 norm. However, this term is sensitive to outliers, and enforces smoothness of the solution, which is not always desirable with trench structures, which typically have a non-smooth profile. The second term is a sparsity regulariser, promoting a piecewise constant solution \cite{Gholami:2010}. Since we will model trenches with soft or steep slopes, we include both terms and choose $\la_1 = \la_2 = \la$.  We manually chose an optimal value of $\la$ from studying a synthetic data set with given signal-to-noise ratio. 

To avoid unphysical solutions, we restrict the conductivity values to lie between given bounds, i.e. 
$\boldsymbol{\s}_- \leq \boldsymbol{\s} \leq \boldsymbol{\s}_+$, and we allow only non-negative layer thickness.

\subsubsection{Implementation}

The inversion procedure was implemented in Matlab \cite{matlab}. Minimisation of the cost function was conducted using the built-in routine \verb!fmincon.m! for constrained non-linear minimisation. The sequential quadrature programming (sqp) algorithm \cite{Nocedal:1991} was selected as it converged faster than the other routines. The Tikhonov regularisation parameter was set manually to $\lambda = 0.2$. We applied the identical regularisation and constraints to both the McNeill forward model \eqref{MNfwd} and 2D forward model \eqref{2Dfwd}. 

The minimisation process of a profile typically requires 100-1000 iteration steps; in each iteration step the forward model \eqref{2Dfwd} is solved hundreds of times, which motivates a fast implementation of the cost function using look-up tables. We create look-up tables before the start of inversion, at the time when the measurement points $x_m$ and reconstruction points $x_r$ are known. In this study, we only invert data along a single measurement line. 

The time-consuming step in the evaluation of $\hat{\s}_{app}(x_m)$ by Eq. \eqref{2Dfwd} are the integrals
\begin{align}
 I_m = \int_{-\infty}^{+ \infty} d\tilde{x} R^{2D}(\tilde{x}, \tilde{z}(x_m + \tilde{x})). 
\end{align}
where we have omitted the layer-index $j$ for brevity. We discretise this expression on the grid of reconstruction points $x_r$ with spacing $\Delta_r = \tilde{x}_r+1 - \tilde{x}_r $:
\begin{align}
 I_m &\approx 
\sum_r \int_{\tilde{x}_r - \Delta_r/2}^{\tilde{x}_{r}+\Delta_r/2} R^{2D}(\tilde{x}_r - \tilde{x}_m, \tilde{z}) \Delta_r = \sum_r C_{mr}  \label{Cmn}
\end{align}
%
%
%
%
We cannot use simple trapezoidal (or quadratic) integration here, since for sparse sampling of the reconstruction points, the significant contribution of $R^{2D}$ may lie between subsequent evaluation points of the integrand. 
Instead, we locally approximate $\tilde{z}$ as piecewise constant on intervals of length $\Delta_r$ centred around the reconstruction points $x_r$ to find Eq. \eqref{Cmn}. In essence, this is a finite-element approach with piecewise constant basis functions. 

The coefficients $C_{mr}$ still depend on the local interface depth $\tilde{z}(\tilde{x}_r)$, which is updated during inversion. Therefore, we precompute a discrete set of lookup values $\tilde{z}_k$ spaced $0.05s $ up to a depth of $4s$:
\begin{align}
  L_{mrk} = \int_{\tilde{x}_r - {\Delta}_r/2}^{\tilde{x}_{r}+{\Delta}_r/2} R^{2D}(\tilde{x} - \tilde{x}_m, \tilde{z}_k) d\tilde{x}
\end{align}
For intermediate z-values, we compute the linear interpolation coefficients $\tilde{z}_k = \lfloor  \tilde{z} \rfloor$ and $f_k = ( \tilde{z} -  \tilde{z}_k)/( \tilde{z}_{k+1} -  \tilde{z}_k)$, such that $I_m$ can be accessed as a linear combination of lookup values:
\begin{align}
  C_{mr} = \sum_r f_{k,r} L_{m,r,k}  +  \sum_r (1-f_{k,r}) L_{m,r,k+1}.
\end{align}
This way, the full 2D forward model is formulated as a linear combination (in x) and linear interpolation (in z) of tabulated values. 

\subsection{Numerical modelling of a trench}

\subsubsection{Trench geometry}
We tested our inversion algorithm on a synthetic dataset. The ambient soil has $\s_1 = 6\,$mS/m and the trench is filled with material that has $\s_2 = 12\,$mS/m. The trench profile is given by superimposing two ramp profiles with transition width $d$ around position $x_0$: 
\begin{align}
   z &= z_0 [ H(-x+x_0+w/2; d) + H(x-x_0-w/2; d)] \nn \\
   H(x;d) &= \frac{1}{2} ( 1 + \tanh(x/d)) \label{trench}
\end{align}
In the limit of $d \rightarrow 0$, $H(x;d)$ becomes a Heaviside function and a rectangular cross-section is obtained of width $w$ and depth $z_0$. For finite $d$, the trench depth $z_{max}$ can be determined from \eqref{trench}. 

To the forward model, we added white Gaussian noise of either $30$ dB (in power units) to the measured signal. 

In a transect of 10m long, we took  $N_m = 101$ measurement points with spacing 10cm. We bisected each interval to get $N_r = 201$ points where $z(x)$ will be reconstructed, i.e. with spatial resolution $5cm$. This approach mimic the limited sampling that may also be present in actual surveys. In our present calculation, all measurement points also happened to be reconstruction points, but this is no constraint on the method. To avoid a possible bias due to choosing the deepest point of the trench on a grid point, we choose the trench midpoint $x_0$ to take a random value between $0$ and $0.1$m. 

\subsubsection{Post-processing}

To assess the quality of reconstruction, we define $x_{50,L}$ and $x_{50,R}$ as the horizontal positions to the left and right of the deepest point which satisfy $z(x_{50}) = z_{max}/2$. We define the trench width to be the full width at half maximum (FWHM): $\hat{w} = x_{50,R} - x_{50,L}$ and the middle of the trench to be $\hat{x}_0 = (x_{50,R} + x_{50,L})/2$. 

\section{Results \label{sec:inversion}}

\subsection{Trench shape}

\begin{figure*}[t]
\vci{\begin{tabular}{ll}
&a) w=0.5 m \\
& \includegraphics[width = 0.4 \textwidth]{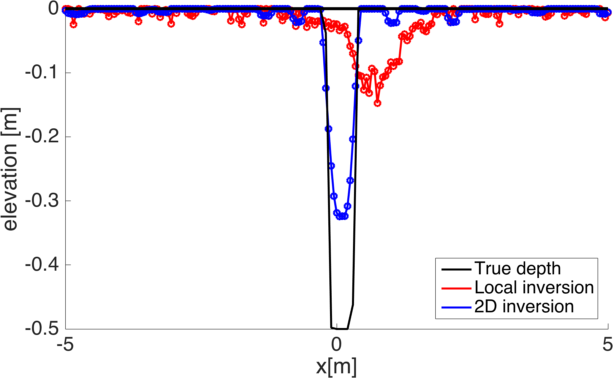}
\end{tabular}}
\vci{
\includegraphics[width = 0.6 \textwidth]{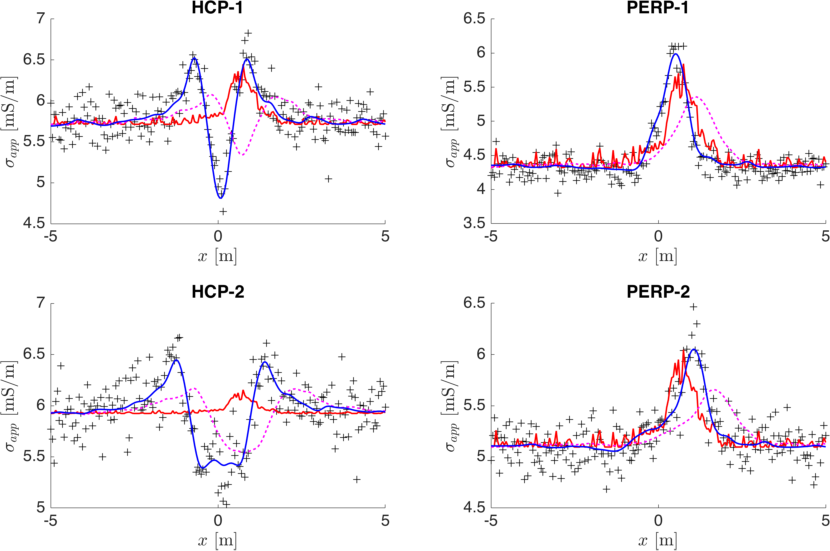}
}\\
\vci{\begin{tabular}{ll}
b) &w=1.5 m \\
& \includegraphics[width = 0.4 \textwidth]{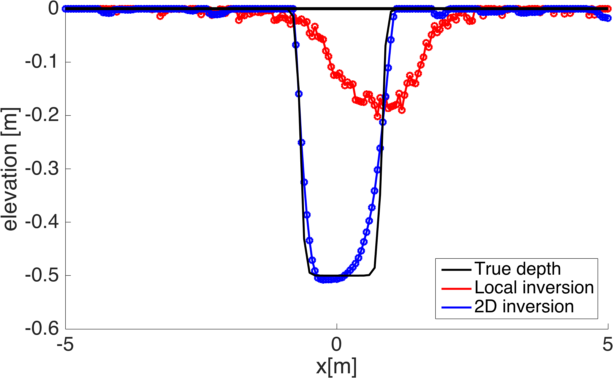}
\end{tabular}}
\vci{
\includegraphics[width = 0.5 \textwidth]{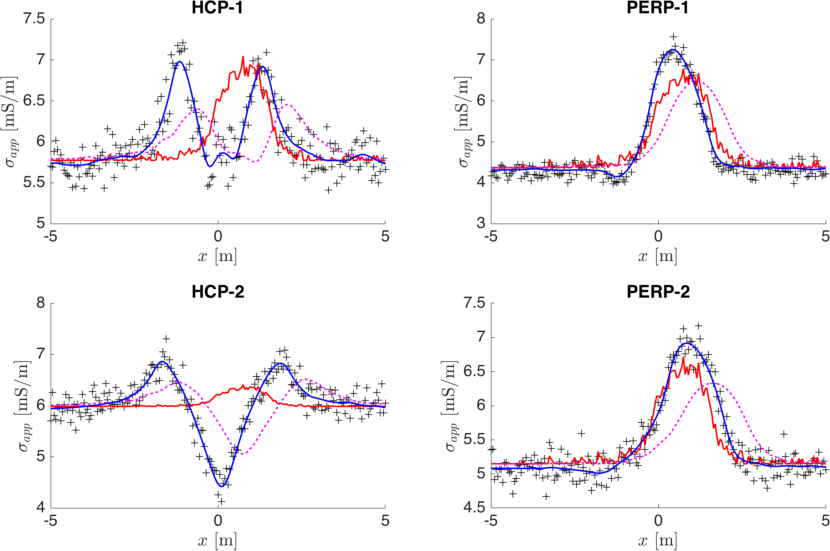}
} \\
\vci{\begin{tabular}{ll}
c) &w=3 m \\
& \includegraphics[width = 0.4 \textwidth]{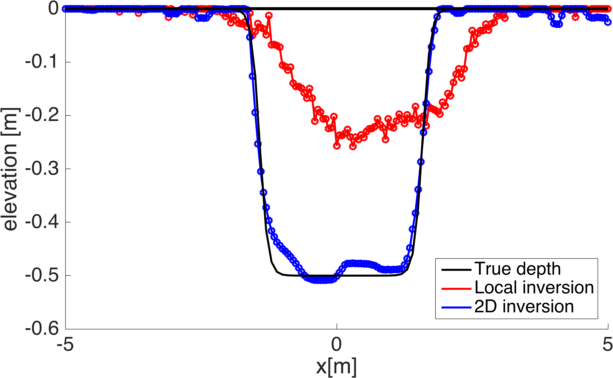}
\end{tabular}}
\vci{
\includegraphics[width = 0.5 \textwidth]{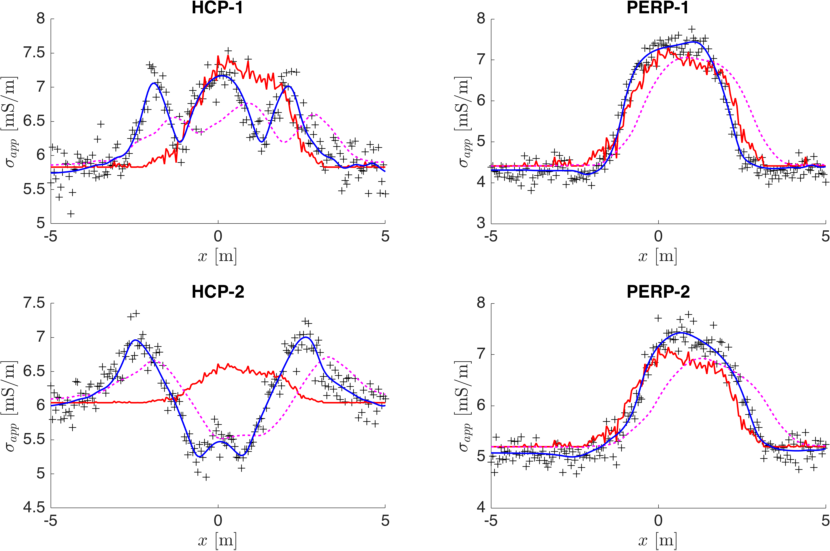}
} 
\caption{\label{fig:shallowsteep} Inversion of simulated EMI-data: shallow trench with steep sides: $z_0 = 0.5$\,m and $d/w=0.05$. SNR = 30, $\lambda$ = 0.02. Rightmost panels show the observed data (+), as well as reconstructed data based on the profiles found by inversion (left panel), for the 1D (red) and 2D (blue) forward model. The magenta line shows the 2D forward model applied to the reconstructed profile using 1D inversion. 
} 
\end{figure*}

\begin{figure*}[t]
\vci{\begin{tabular}{ll}
&a) w=0.5 m \\
& \includegraphics[width = 0.4 \textwidth]{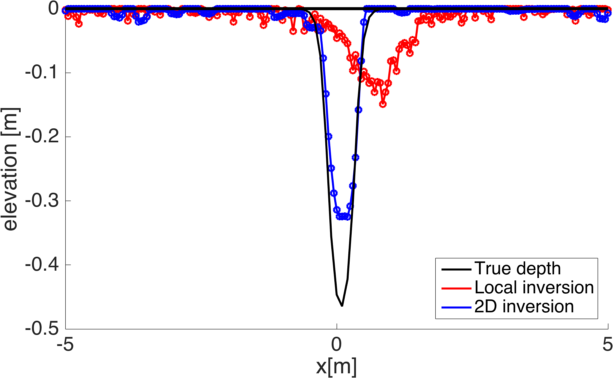}
\end{tabular}}
\vci{
\includegraphics[width = 0.6 \textwidth]{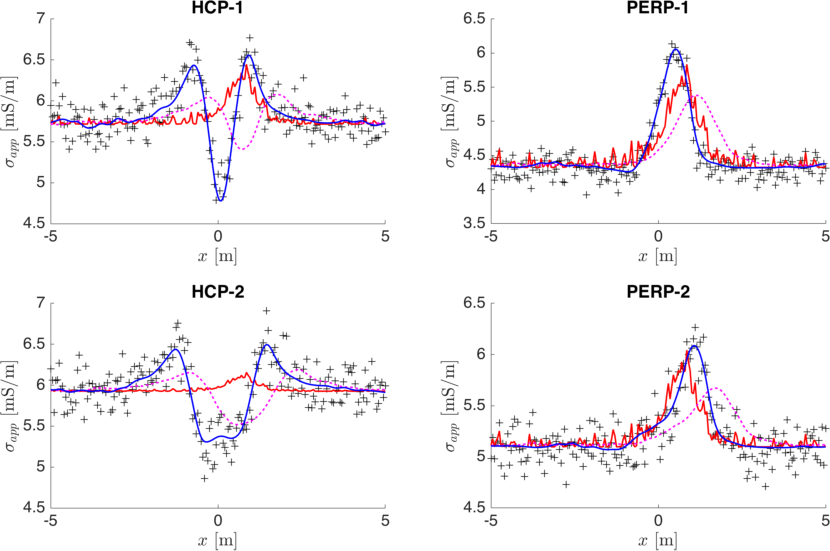}
}\\
\vci{\begin{tabular}{ll}
b) &w=1.5 m \\
& \includegraphics[width = 0.4 \textwidth]{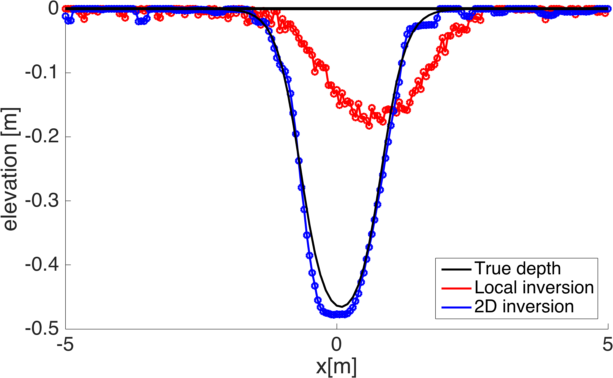}
\end{tabular}}
\vci{
\includegraphics[width = 0.5 \textwidth]{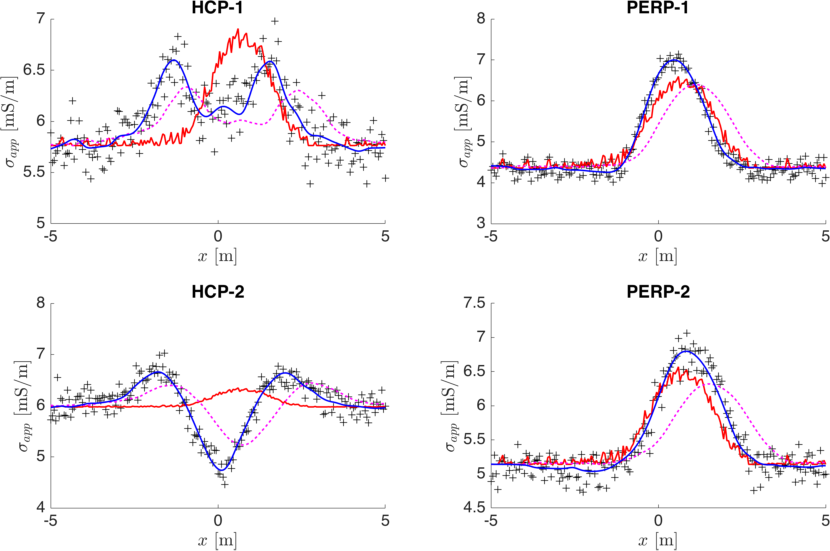}
} \\
\vci{\begin{tabular}{ll}
c) &w=3 m \\
& \includegraphics[width = 0.4 \textwidth]{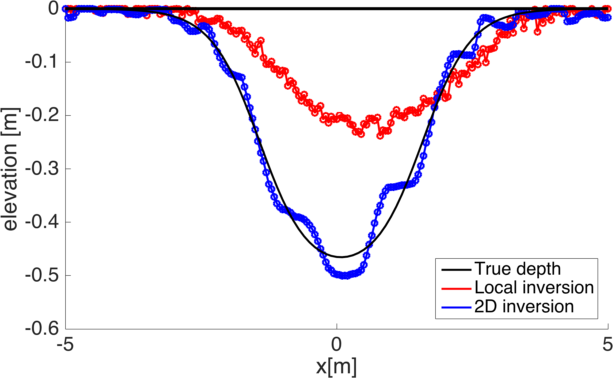}
\end{tabular}}
\vci{
\includegraphics[width = 0.5 \textwidth]{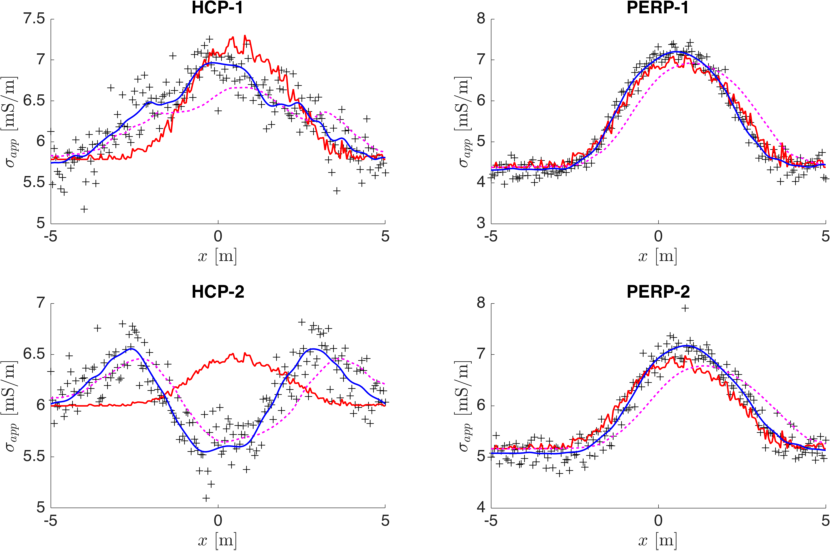}
} 
\caption{\label{fig:shallownonsteep} Inversion of simulated EMI-data: shallow trench with non-steep sides: $z_0 = 0.5$\,m and $d/w=0.3$. SNR = 30, $\lambda$ = 0.02.} 
\end{figure*}

\begin{figure*}[t]
\vci{\begin{tabular}{ll}
&a) w=0.5 m \\
& \includegraphics[width = 0.4 \textwidth]{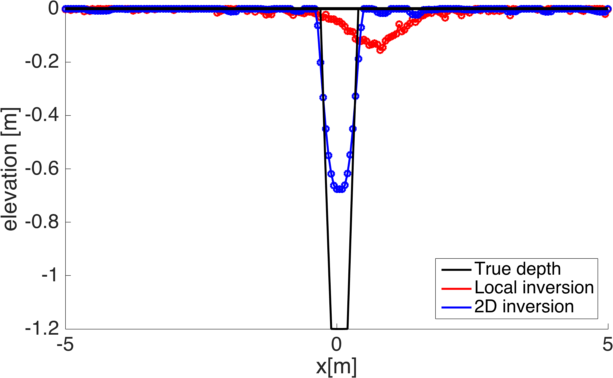}
\end{tabular}}
\vci{
\includegraphics[width = 0.6 \textwidth]{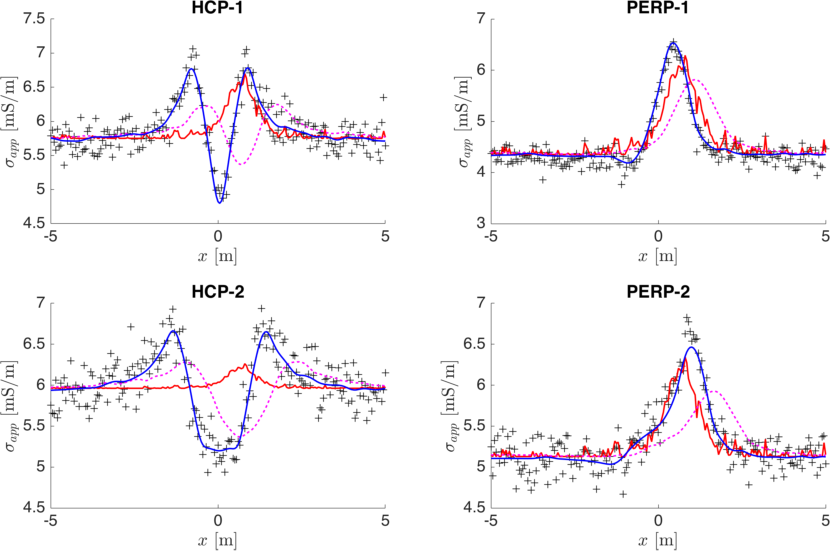}
}\\
\vci{\begin{tabular}{ll}
b) &w=1.5 m \\
& \includegraphics[width = 0.4 \textwidth]{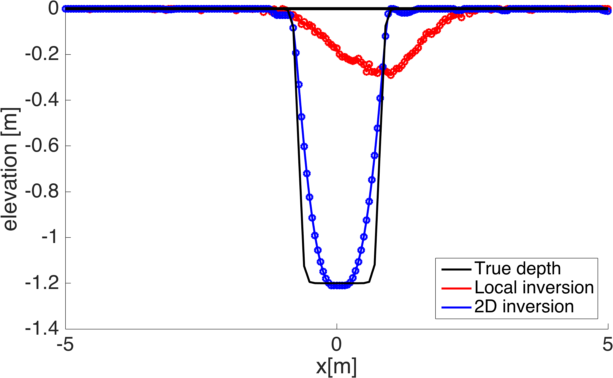}
\end{tabular}}
\vci{
\includegraphics[width = 0.5 \textwidth]{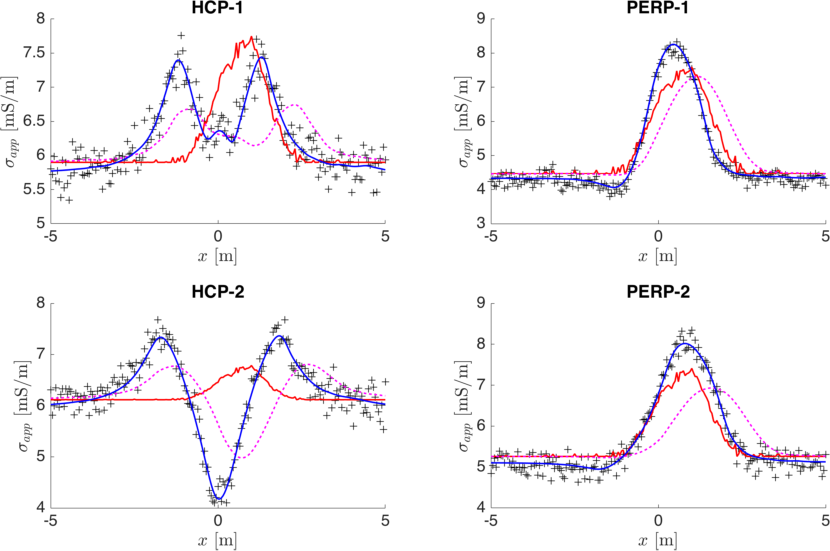}
} \\
\vci{\begin{tabular}{ll}
c) &w=3 m \\
& \includegraphics[width = 0.4 \textwidth]{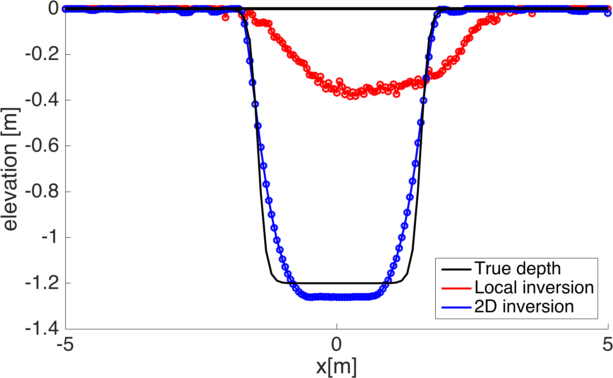}
\end{tabular}}
\vci{
\includegraphics[width = 0.5 \textwidth]{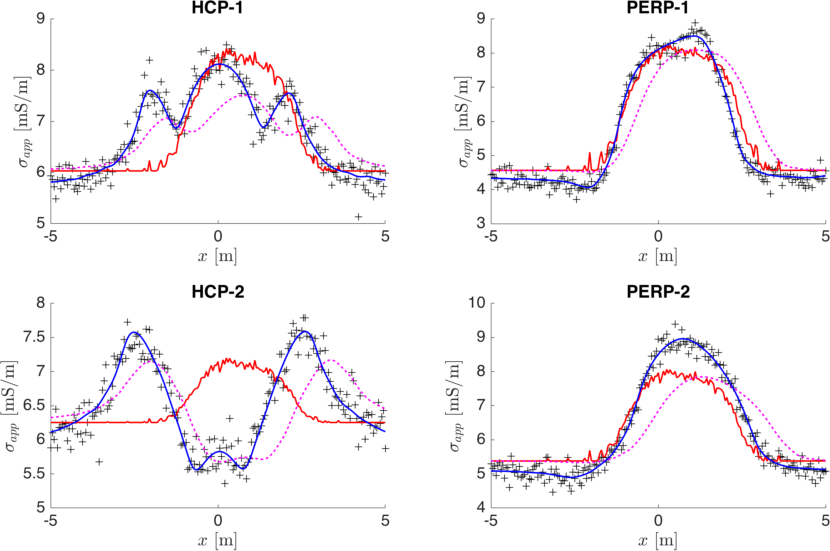}
} 
\caption{\label{fig:deepsteep} Inversion of simulated EMI-data: deep trench with steep sides: $z_0 = 1.2$\,m and $d/w=0.05$. SNR = 30, $\lambda$ = 0.02.} 
\end{figure*}

\begin{figure*}[t]
\vci{\begin{tabular}{ll}
&a) w=0.5 m \\
& \includegraphics[width = 0.4 \textwidth]{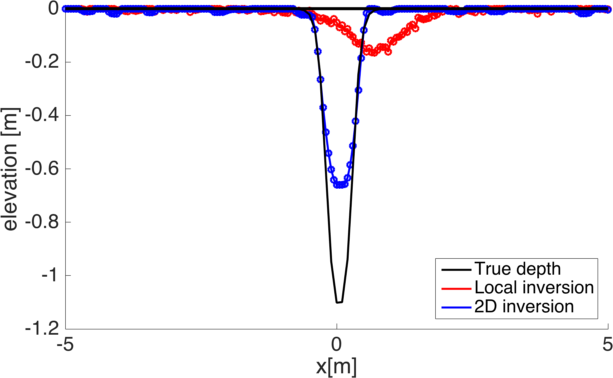}
\end{tabular}}
\vci{
\includegraphics[width = 0.6 \textwidth]{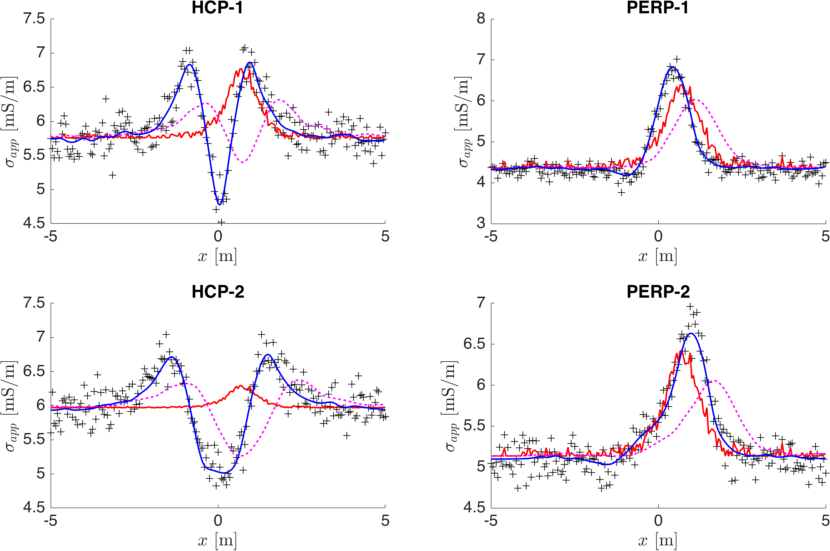}
}\\
\vci{\begin{tabular}{ll}
b) &w=1.5 m \\
& \includegraphics[width = 0.4 \textwidth]{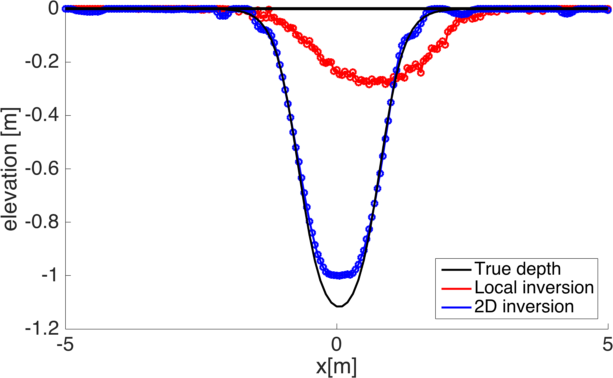}
\end{tabular}}
\vci{
\includegraphics[width = 0.5 \textwidth]{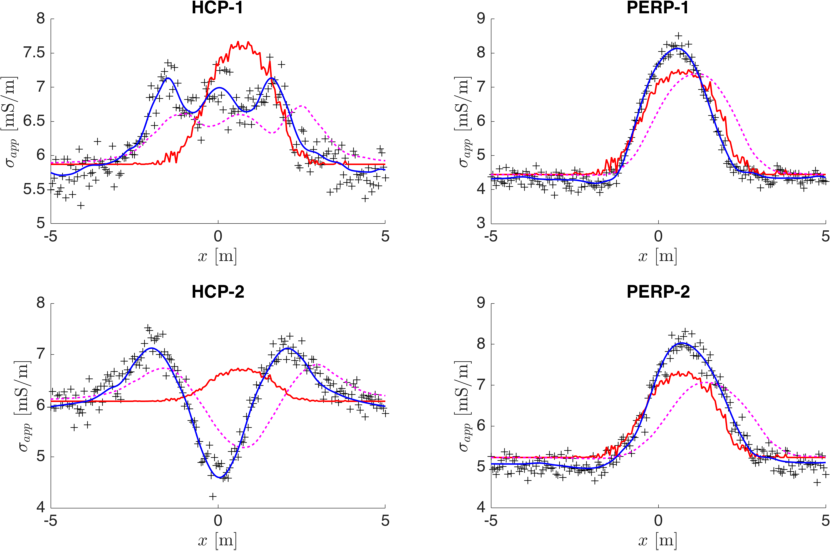}
} \\
\vci{\begin{tabular}{ll}
c) &w=3 m \\
& \includegraphics[width = 0.4 \textwidth]{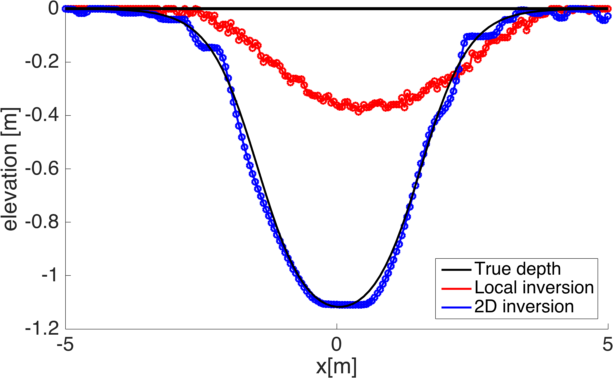}
\end{tabular}}
\vci{
\includegraphics[width = 0.5 \textwidth]{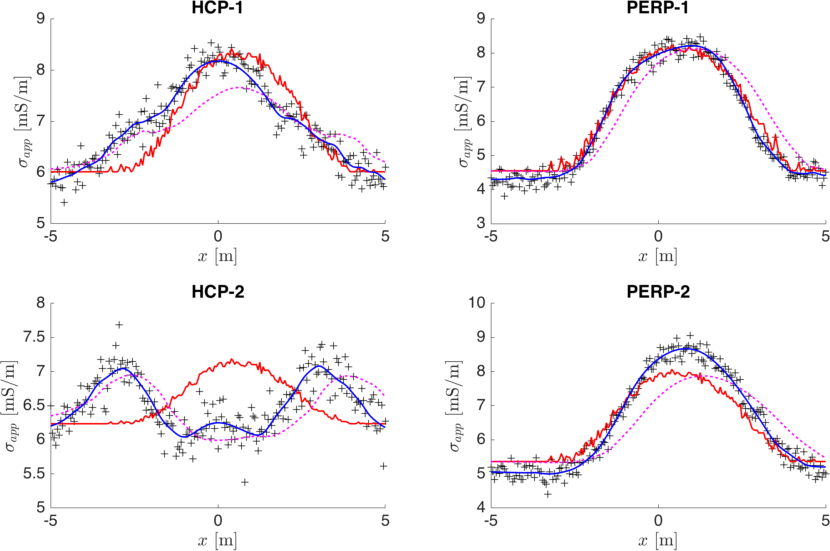}
} 
\caption{\label{fig:deepnonsteep} Inversion of simulated EMI-data: deep trench with non-steep sides: $z_0 = 1.2$\,m and $d/w=0.3$. SNR = 30, $\lambda$ = 0.02.} 
\end{figure*}

Figures \ref{fig:shallowsteep} - \ref{fig:deepnonsteep} show the results of inversion, for various trench parameters, including widths $w$, depths $z$ and steepness $d$. On the left, the ground truth $z(x)$ is presented, as well as results from 1D (red) and 2D (blue) reconstruction. The 1D and 2D reconstructions only differ in the forward model used; identical lateral regularization was imposed. The rightmost panels of these figures show the observed $\s_{app}$ (black $+$) for each of the 4 receiver coils of a Dualem 21 instrument. On these data, we superimpose the reconstructed data of the 2D forward model (blue) and 1D forward model (red). An additional curve (magenta) is given for the 1D reconstruction fed into the 2D forward model, showing that the 1D method cannot correctly model sudden horizontal transitions in the soil conductivity. 

We show 12 cases in total, in 4 figures where different trench widths are displayed (0.5, 1.5 and 3 \, m). The figures correspond to shallow trenches (Fig.  \ref{fig:shallowsteep} \&  \ref{fig:shallownonsteep} ) and deep trenches (Fig.  \ref{fig:deepsteep}  \&  \ref{fig:deepnonsteep} ), which can have either gradual or steep slopes. 

From the reconstructed dataset, it can be seen that the 2D model better covers the W-shape in the HCP coils, 
while the 1D inversion only covers the monomodal course of the PERP coils. Note that despite the significant level of added noise, the inversion method can still distill a coherent profile. 

Furthermore, we see that choosing a single fixed value for $\lambda_1$ (promoting sudden changes) and $\lambda_2$ (promoting gradual changes in interface depth) may not be optimal, since non-steep slopes may experience a staircase artefact (see e.g. \eqref{fig:shallowsteep}. Thus, if a priori information is available on the trench shape, it can be used to set the regularisation parameters. However, to show generic results of this method, we refrain from case-based tuning of $\lambda$ in this study. 

For intermediate cases in width and steepness, the 2D inversion result recovers well the trench position, depth  and width. 

%

\subsection{Estimation of trench parameters}

\begin{figure}[b] \centering
\begin{tabular}{c @{\hspace{5cm}} c}
  SNR = 30 & SNR= 50 
\end{tabular}\\ 
\raisebox{4cm}{a)}
 \includegraphics[width = 0.43 \textwidth]{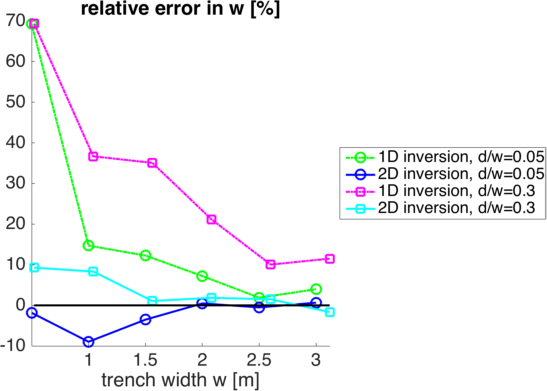}
 \includegraphics[width = 0.43 \textwidth]{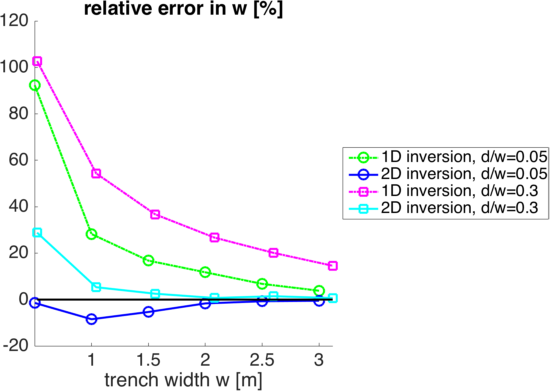}\\
\raisebox{4cm}{b)}
 \includegraphics[width = 0.43 \textwidth]{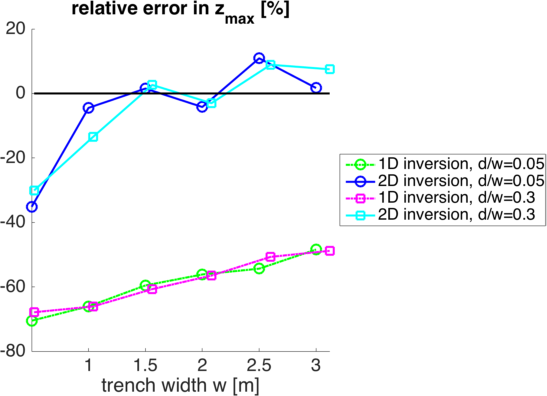}
 \includegraphics[width = 0.43 \textwidth]{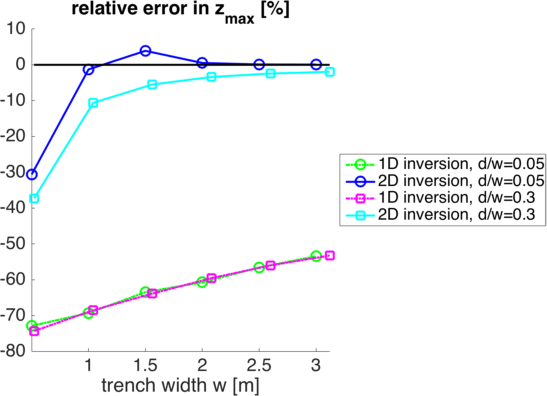}\\
\raisebox{4cm}{c)}
 \includegraphics[width = 0.43 \textwidth]{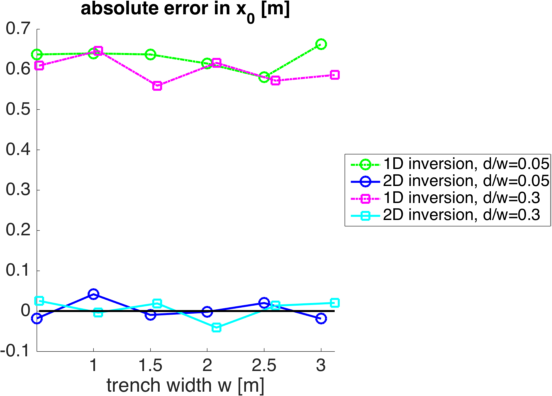}
 \includegraphics[width = 0.43 \textwidth]{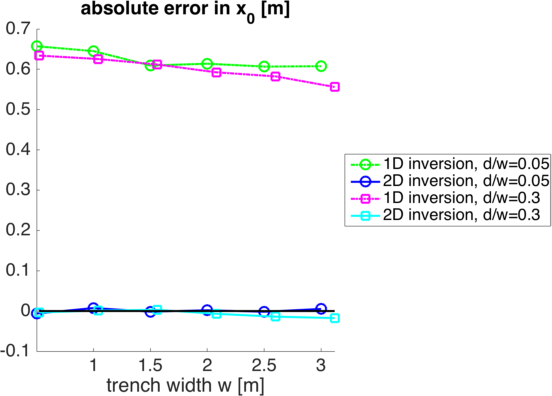}\\
\caption{\label{fig:performance_0,5} Evaluation of inversion quality by comparing the reconstructed values of trench width $w$, depth $z_{max}$ and  position $x_0$ with ground-truth. Absolute errors of inversion results for a trench of nominal depth $z_0 = 0.5$\,m: dependency on trench width $w$, slope parameter $d/w$ and SNR (left column: SNR=30, right column: SNR = 50). Both 1D and 2D inversion results used the same regularisation parameter $\lambda = 0.02$. \label{fig:stats_d=0,5}
}
\end{figure}

\begin{figure}[b] \centering
\begin{tabular}{c @{\hspace{5cm}} c}
  SNR = 30 & SNR= 50 
\end{tabular}\\ 
\raisebox{4cm}{a)}
 \includegraphics[width = 0.43 \textwidth]{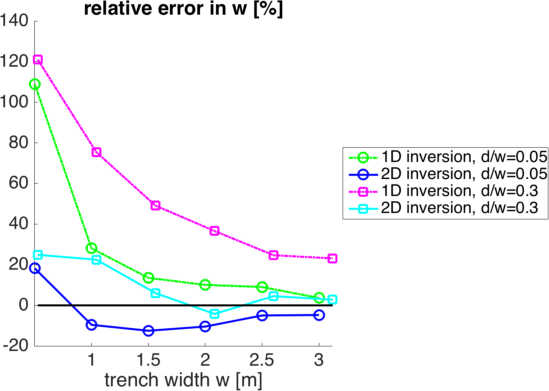}
 \includegraphics[width = 0.43 \textwidth]{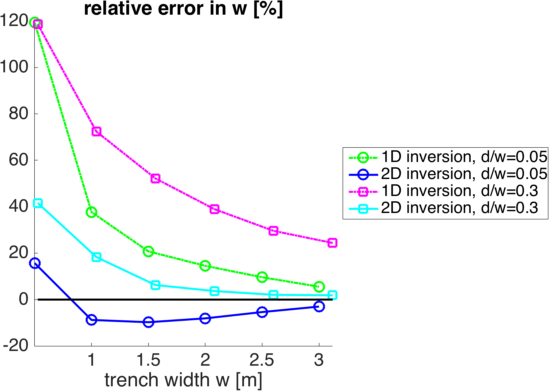}\\
\raisebox{4cm}{b)}
 \includegraphics[width = 0.43 \textwidth]{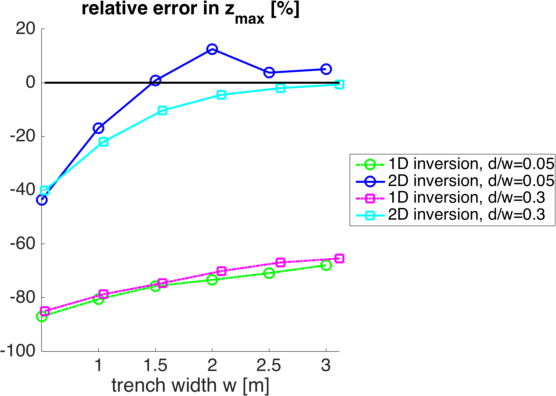}
 \includegraphics[width = 0.43 \textwidth]{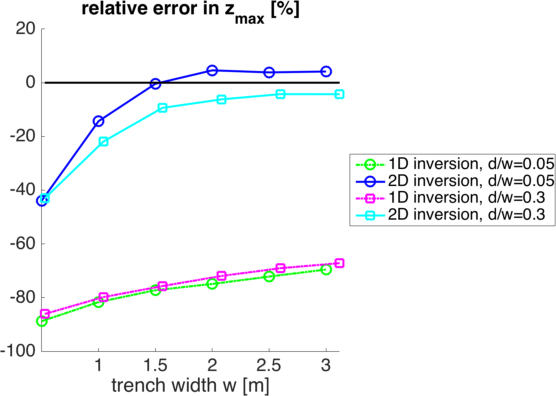}\\
\raisebox{4cm}{c)}
 \includegraphics[width = 0.43 \textwidth]{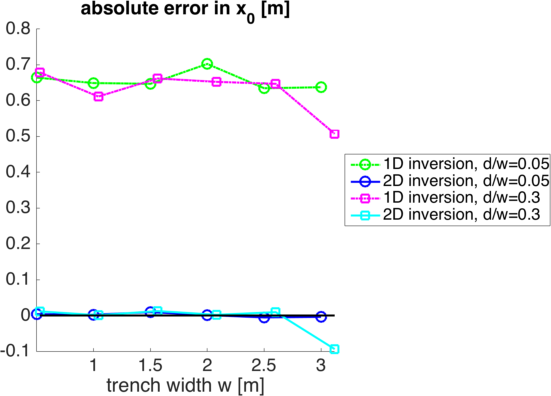}
 \includegraphics[width = 0.43 \textwidth]{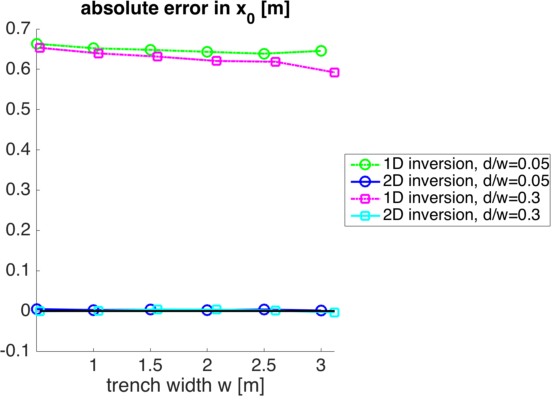}\\
\caption{\label{fig:performance_1,2} Evaluation of inversion quality by comparing the reconstructed values of trench width $w$, depth $z_{max}$ and  position $x_0$ with ground-truth. Same as Fig. \ref{fig:stats_d=0,5} for a trench depth of $z_0 = 1.2$m. \label{fig:stats_d=1,2}
}
\end{figure}

\clearpage
\section{Discussion \label{sec:discussion}}

In this work we have developed a 2D inversion method for electromagnetic induction survey data. The method was tested on synthetic data. In the range where a 2D inversion may be feasible, i.e. sudden horizontal transitions in conductivity on the scale of the intercoil distance, the 2D method is seen to outperform the 1D method. The latter suffers from systematic error in the determination of the PERP peak, which is reflected in a constant error in the estimation of the trench position. 
The 1D model moreover neglects the typical W-shape of the HCP readings associated to a trench structure, therefore it fails to reproduce the complexity of data generated with a 2D model. As such, trench depth is largely underestimated by the 1D model, with errors in the range of 50-80\%. 
In contract, the 2D model can estimate trench parameters within few \% error for trench widths above 2m. Our method can also handle narrow trenches ($w \leq s$), as long as the spatial sampling is dense enough to provide at least 5 data points in both transition zones. 

An open question remains choosing an appropriate regularisation parameter $\lambda$. Here, we took $\lambda$ to be the smallest value that did not deliver obvious visual artefacts (spikes, staircase effects) in the reconstruction in the entire range of trench widths used. It can be expected that in practice, $\la$ has to be adapted to the SNR of the data, and thereby the final reconstruction will reflect the quality of the collected data. 

Our method is currently tailored the case where the direction of scanning is perpendicular to the alignment of the trench. If desired, this case can be easily realised on the terrain, by a first quick survey to find the orientation of 1D structures, followed by a detailed scan transverse to the linear structure to be observed. However, the expressions \eqref{phi3} are general and integrals thereof may also be tabulated for other intersection angles between trench and scanning direction. However, in this case, it may be more practical to develop a true 3D inversion method, which for a layered earth may also be analytically feasible, complementing 3D meshing approaches currently taken by other groups \cite{sasaki:2001, sasaki:2010, noh:2016}. 

In this work, we have presented a 2D inversion method for geophysical survey data based on a physical model. Our tests on \textit{in silico} data reveal that the method may outperform inversion methods based on regularisation of an 1D forward model. 


\bibliographystyle{unsrt}

\begin{thebibliography}{10}

\bibitem{huang:2017}
J.~Huang, A.~Pedrera-Parrilla, K.~Vanderlinden, E.V. Taguas, J.A. Gómez, and
  J.~Triantafilis.
\newblock Potential to map depth-specific soil organic matter content across an
  olive grove using quasi-2d and quasi-3d inversion of {DUALEM}-21 data.
\newblock {\em Catena}, 152:207--217, May 2017.

\bibitem{dakak:2017}
H.~Dakak, J.~Huang, A.~Zouahri, A.~Douaik, and J.~Triantafilis.
\newblock Mapping soil salinity in 3-dimensions using an {EM}38 and {EM}4soil
  inversion modelling at the reconnaissance scale in central {Morocco}.
\newblock {\em Soil Use and Management}, 33(4):553--567, December 2017.

\bibitem{Grzegorczyk:2012}
Tomasz~M. Grzegorczyk, Juan~Pablo Fernández, Fridon Shubitidze, Kevin O'Neill,
  and Benjamin~E. Barrowes.
\newblock Subsurface electromagnetic induction imaging for unexploded ordnance
  detection.
\newblock {\em Journal of Applied Geophysics}, 79:38--45, April 2012.

\bibitem{oneill:2016}
K.~O'Neill.
\newblock {\em Discrimination of Subsurface Unexploded Ordnance}.
\newblock SPIE books.

\bibitem{Vallee:2011}
Smith~R. Vall\'ee, M. and P.~Keating.
\newblock Metalliferous mining geophysics —state of the art after a decade in
  the new millennium.
\newblock 2011.

\bibitem{Beard:1998}
Les~P. Beard and Jonathan~E. Nyquist.
\newblock Simultaneous inversion of airborne electromagnetic data for
  resistivity and magnetic permeability.
\newblock {\em Geophysics}, 63(5):1556--1564, 1998.

\bibitem{DeSmedt:2014}
Philippe De~Smedt, Marc Van~Meirvenne, Timothy Saey, Eamonn Baldwin, Chris
  Gaffney, and Vince Gaffney.
\newblock Unveiling the prehistoric landscape at stonehenge through
  multi-receiver emi.
\newblock {\em Journal of Archaeological Science}, 50:16--23, 2014.

\bibitem{Saey:2014}
Timothy Saey, Samuël Delefortrie, Lieven Verdonck, Philippe De~Smedt, and Marc
  Van~Meirvenne.
\newblock Integrating {EMI} and {GPR} data to enhance the three-dimensional
  reconstruction of a circular ditch system.
\newblock {\em Journal of Applied Geophysics}, 101:42--50, February 2014.

\bibitem{Wait:1962}
James~R. Wait.
\newblock A note on the electromagnetic response of a stratified earth.
\newblock {\em Geophysics}, 27(3):382--385, 1962.

\bibitem{McNeill:1980}
J.D. McNeill.
\newblock Electromagnetic terrrain conductivity measurement at low induction
  numbers.
\newblock Technical report, Geonics Ltd., 1980.

\bibitem{Delrue:2018}
S.~Delrue, D.~Dudal, and B.~Maveau.
\newblock A damped forward emi model for a horizontally stratified earth.
\newblock 2018.

\bibitem{Tikhonov:1977}
A.~N. Tikhonov and V.~Y. Arsenin.
\newblock {\em Solution of Ill-posed Problems}.
\newblock Winston \& Sons, Washington, 1977.

\bibitem{auken:2004}
E.~Auken and A.~V. Christiansen.
\newblock Layered and laterally constrained 2d inversion of resistivity data.
\newblock 69:752, 2004.

\bibitem{MonteiroSantos:2004}
Fernando~A Monteiro~Santos.
\newblock 1-{D} laterally constrained inversion of {EM}34 profiling data.
\newblock {\em Journal of Applied Geophysics}, 56(2):123--134, June 2004.

\bibitem{MonteiroSantos:2010}
F.~A.~Monteiro Santos, J.~Triantafilis, K.~E. Bruzgulis, and J.~A.~E. Roe.
\newblock Inversion of {Multiconfiguration} {Electromagnetic} ({DUALEM}-421)
  {Profiling} {Data} {Using} a {One}-{Dimensional} {Laterally} {Constrained}
  {Algorithm}.
\newblock {\em Vadose Zone Journal}, 9(1):117, 2010.

\bibitem{farquharson:2003}
Colin~G. Farquharson, Douglas~W. Oldenburg, and Partha~S. Routh.
\newblock Simultaneous 1d inversion of loop–loop electromagnetic data for
  magnetic susceptibility and electrical conductivity.
\newblock {\em Goephysics}, 68(6):1857--1869, November 2003.

\bibitem{brodie:2009}
R.~Brodie and M.~Sambridge.
\newblock Holistic inversion of frequencydomain airborne electromagnetic data
  with minimal prior information.
\newblock 2009.

\bibitem{Gholami:2010}
A.~Gholami and H.~R. Siahkoohi.
\newblock Regularization of linear and non-linear geophysical ill-posed
  problems with joint sparsity constraints.
\newblock {\em Geophysical Journal International}, 180(2):871--882, 2010.

\bibitem{sasaki:2001}
Y.~Sasaki.
\newblock Full 3-d inversion of electromagnetic data on a {PC}.
\newblock {\em Journal of Applied Geophysics}, 46:45, 2001.

\bibitem{sasaki:2010}
Yutaka Sasaki, Jung-Ho Kim, and Seong-Jun Cho.
\newblock Multidimensional inversion of loop-loop frequency-domain {EM} data
  for resistivity and magnetic susceptibility.
\newblock {\em Geophysics}, 75(6):F213--F223, November 2010.

\bibitem{noh:2016}
Kyubo Noh, Seokmin Oh, Soon~Jee Seol, Ki~Ha Lee, and Joongmoo Byun.
\newblock Analysis of anomalous electrical conductivity and magnetic
  permeability effects using a frequency domain controlled-source
  electromagnetic method.
\newblock {\em Geophysical Journal International}, 204(3):1550--1564, March
  2016.

\bibitem{thiesson:2017}
J.~Thiesson, A.~Tabbagh, F.-X. Simon, and M.~Dabas.
\newblock 3d linear inversion of magnetic susceptibility data acquired by
  frequency domain {EMI}.
\newblock {\em Journal of Applied Geophysics}, 136:165--177, January 2017.

\bibitem{Jackson:1999}
John~David Jackson.
\newblock {\em Classical electrodynamics}.
\newblock Wiley, New York, {NY}, 3rd ed. edition, 1999.

\bibitem{Gradstein:1965}
I.S.Gradstein and I.M.Ryzhik.
\newblock {\em Tables of Integrals, Sums, Series, and Products}.
\newblock Acad. Press, New York, 1965.

\bibitem{matlab}
Matlab and statistics toolbox release 2012b, the mathworks, inc., {N}atick,
  {M}assachusetts, {United States}.

\bibitem{Nocedal:1991}
J.~Nocedal and S.~J. Wright.
\newblock {\em Numerical Optimization}.
\newblock 2006.

\end{thebibliography}

\end{document}